\documentstyle[a4,12pt,epsfig]{article}
\def\BaBar{\mbox{$\mathrm{Ba\overline Bar} \:$}}

\def\b{\mbox{$b$}}

\def\cbarc{\mbox{$c\bar c$}}

\def\Vudstar{\mbox{$V^{*}_{ud}$}}

\def\Vub{\mbox{$V_{ub}$}}

\def\Vtdstar{\mbox{$V^{*}_{td}$}}

\def\Vtb{\mbox{$V_{tb}$}}

\def\sen2alpha{\mbox{$\sin 2 \alpha \,$}}
\def\sen2beta{\mbox{$\sin 2 \beta \,$}}
\def\btojpsi{\mbox{$ {B^{\scriptscriptstyle 0}_{\scriptscriptstyle d}}
     \to J / \psi K^{\scriptscriptstyle 0}_{S} $}}
\def\bbartojpsi{\mbox{$ \stackrel{{\footnotesize (-)}}
     {B^{\scriptscriptstyle 0}_{\scriptscriptstyle d}}
     \to J / \psi K^{\scriptscriptstyle 0}_{S} $}}
\def\b&bartopipi{\mbox{$\overline {B^{0}} (B^{0}) \to {\pi^{+}} {\pi^{-}} $}}
\def\bbartopipi{\mbox{$\stackrel{{\scriptscriptstyle (-)}}
    {B^{\scriptscriptstyle 0}_{\scriptscriptstyle d}} 
    \to {\pi^{+}} {\pi^{-}} $}}
\def\btopipi{\mbox{$ {B^{\scriptscriptstyle 0}_{\scriptscriptstyle d}}
     \to {\pi^{+}} {\pi^{-}} \/ $}}
\def\pipi{\mbox{$ {\pi^{+}} {\pi^{-}} \/ $}}
\def\btopp{\mbox{$ {B^{\scriptscriptstyle 0}_{\scriptscriptstyle d}}
     \to \pi \pi \/ $}}

\def\btopi0pi0{\mbox{$ {B^{ \scriptscriptstyle 0}_{\scriptscriptstyle d}} 
     \to \pi^{\scriptscriptstyle 0} \pi^{\scriptscriptstyle 0} \/  $}} 
\def\btok0k0bar{\mbox{$ {B^{ \scriptscriptstyle 0}_{\scriptscriptstyle d}} 
     \to K^{\scriptscriptstyle 0} 
     \overline {K^{\scriptscriptstyle 0}} \/  $}} 

\def\btoduu{\mbox{$ b\to ud \overline {u} $}}
\def\alpham{\mbox{${\alpha_{\scriptscriptstyle M}}$}}
\def\Ap{\mbox{${A_{\scriptscriptstyle P}}$}}
\def\At{\mbox{${A_{\scriptscriptstyle T}}$}}
\def\Apsuat{\mbox{${A_{\scriptscriptstyle P}} / {A_{\scriptscriptstyle T}} $}}
\def\phit{\mbox{${\phi_{\scriptscriptstyle T}}$}}
\def\phip{\mbox{${\phi_{\scriptscriptstyle P}}$}}
\def\deltat{\mbox{${\delta_{\scriptscriptstyle T}}$}}
\def\deltap{\mbox{${\delta_{\scriptscriptstyle P}}$}}

\def\azero{\mbox{$a_{0} \:$\/}}
\def\azerobar{\mbox{$\overline {a_{0}} \:$}}
\def\bzero{\mbox{$b_{0} \:$\/}}
\def\bzeromax{\mbox{${b^{\:max}_{0}}$}}

\def\siga0{\mbox{${\sigma}_{\scriptscriptstyle a_{0}} \:$\/}}

\def\sigb0{\mbox{${\sigma}_{\scriptscriptstyle b_{0}} \:$\/}}

\begin{document}
\begin{titlepage}
\begin{flushright} \large
   HEP-PH/9702353
\\
\end{flushright} 
\bigskip \bigskip \bigskip \bigskip
\begin{center} \huge \bf
  \newcommand {\BaBarb} {\boldmath$\BaBar$}
Penguin corrections and strong phases 
in a time-dependent analysis of $\b&bartopipi$ 
\end{center} 
 \bigskip \bigskip
\begin{center} \Large
   P. S.~Marrocchesi  
\end{center}
\begin{center} \large
       (Modena University and INFN/Pisa)
\end{center}
\begin{center} \Large
   N.~Paver 
\end{center}
\begin{center} \large
       (Trieste University and INFN/Trieste)
\end{center} \bigskip \bigskip
\begin{abstract}
 \noindent 
 From a time-dependent analysis of the decay $\bbartopipi$ and 
 using a model-dependent $\Apsuat$ ratio 
 of the penguin-to-tree amplitudes 
 contributing to the decay, {\it both} the weak phase $\alpha$ {\it and}
 the strong  phase shift difference $\delta$ 
 can be extracted from the data.
 The value of the weak phase $\beta$, expected to be  
 measured from the decay $\btojpsi$, is used 
 to parameterize the value of $\Apsuat$ and the 
 corresponding penguin correction to the
 observed asymmetry in $\btopipi$.
\end{abstract}
\bigskip
\end{titlepage}

\section{Introduction}
\label{sec:intro}

 Measurements of CP violation in $\bbartopipi$ and $\bbartojpsi$ decays 
 are expected to provide information, respectively, on the angles $\alpha$ and 
 $\beta$  of the unitarity triangle \cite{bib:quinn}. As it is well known,
 in $\btojpsi$ the weak phase of the penguin
 contribution is the same as in the (largely) dominant tree amplitude 
 and therefore the value of $\sin 2 \beta \: $ can be directly 
 extracted from the measurement of the time-dependent
 CP asymmetry with no hadronic uncertainties. In particular, as being 
 dominated by a single amplitude, 
 this channel should be insensitive to the presence of  
 phases of strong-interaction origin.

 On the other hand, the 
 extraction of $\sin 2\alpha$ from a time-dependent analysis of the asymmetry 
 in $\btopipi$ appears to be less straightforward \cite{bib:gronau93}. 
 This is mainly due to the presence of gluonic and electroweak
 penguin terms with weak phases different from those of the dominant tree 
 $\btoduu$  amplitude. The common prejudice is that the penguin should be 
 smaller than the tree diagram. However, for a quantitatively meaningful 
 statement, the relative size must be either estimated theoretically 
 or measured experimentally from a suitable data sample of $B$-decays.  
 An additional source of uncertainty is introduced by
 strong-interaction phases which can be different for tree and 
 penguin amplitudes and difficult to model at the 
 non-perturbative hadronic level.

 While the size of the tree and penguin amplitudes (in particular, their 
 ratio) can be somehow assessed under some theoretical assumptions,  
 e.g., the factorization hypothesis for non-leptonic decay 
 amplitudes \cite{bib:stech}, 
 the effect of the strong-interaction phases is much harder to estimate 
 reliably \cite{wolfes}. The conventional expectation is that such phases should not be 
 large. This seems verified by the phases of perturbative origin 
 \cite{bib:kramer1},\cite{bib:simma}, coming from the $\cbarc$ penguin with on-shell $\cbarc$ 
 quarks \cite{buras1,bander}, and is 
 expected to be true also for the final state, long distance, 
 non-perturbative phases, considering the high velocity of the produced 
 $\pipi$ pair. Nevertheless, from considerations based on Regge phenomenology, 
 it has been recently suggested that, contrary to expectations, final state 
 interactions might 
 be important even in B non leptonic decays \cite{bib:donoghue}. Also, 
 significant effects have been estimated by semiquantitative (and 
 model-dependent) calculations for the color suppressed channels  
 $B^0\to D^0\pi^0$ and $B^0\to\pi^0\pi^0$ \cite{kamal,halperin}. 

 Several strategies have been proposed to extract the value of 
 $\sin 2\alpha$ in a model-independent way avoiding the difficulty 
 associated with penguins in $\btopipi$.
 However, although quite effective in principle, their
 application to a practical analysis of the data
 might run into some difficulties and the effective potential of these methods
 have still to be realistically assessed by taking into account
 the expected statistics and experimental sensitivities.
 For example, the isospin analysis of $\btopp$ decays \cite{bib:gronau90},
 \cite{xing1},\cite{he} which 
 would enable to directly reconstruct the unitarity triangle, suffers from the 
 predicted low rate for $\btopi0pi0$, of the order of $10^{-6}$ or less 
 \cite{kramer2,deandrea}, and unfavourable background conditions. Alternative 
 methods require more complicated analyses to fit penguin amplitudes by 
 combining measurements of $\btopipi$ with different penguin-dominated 
 processes such as $ B \to K \pi$ \cite{wolfe,xing2,desh1,hernandez,dighe,mannel} or 
 $\btok0k0bar$ \cite{bib:buras2} and relying on SU(3) symmetry of matrix 
 elements and/or first-order SU(3) breaking. In that case, the sensitivity 
 to $\alpha$ could be rather difficult to fully establish in practice. 

 The hadronic uncertainty in the extraction of
 $\sin 2\alpha$ from the CP asymmetry 
 in $\bbartopipi$ has been discussed by many authors
 and the analysis of this channel extended to 
 other, non-strange, $\Delta S=0$ final states such as 
 $\rho \pi$ or $\rho \rho$, and the corresponding $\Delta S = 1$ 
 decays \cite{bib:kramer1,sfigas,snyder,aleksan,fritz}. To derive quantitative 
 results, assumptions are made on the relevant hadronic matrix
 elements (in general, the factorization model), and on the final state 
 strong interaction effects which in most cases are assumed to be negligible 
 .\par 
 In the following, we focus on the determination of $\sin 2\alpha$ from the
 measured CP time 
 asymmetry in $\bbartopipi$. A discussion of the penguin-induced correction 
 $\Delta \alpha$ in terms of $\alpha$ itself and of the strong phase 
 $\delta = \deltat - \deltap$ between tree and penguin amplitudes was 
 presented  
 in \cite{bib:gronau93}, where a first-order expansion of the asymmetry in the 
 penguin-to-tree amplitude ratio $\Apsuat$  was used. As a result of 
 this analysis, 
 it turns out that $\Delta \alpha$ might be significant, except for large 
 values of the asymmetry and that, in any case, the correction 
 should be maximal for $\delta = 0 $.
\footnote{Related analyses can be found, e.g., in \cite{xing3}. }

 The strong phase effect on  $\Delta \alpha$ for general values of $\delta$
 has been further considered in \cite{bib:kramer1} using the exact expressions 
 for the CP asymmetry and the tree-to-penguin amplitude ratio from the 
 factorization model as input. The indication from \cite{bib:kramer1} is 
 that, within the assumed model, the expected relative shift in $\alpha$ should 
 not exceed $30 \%$ 
 and, similarly to \cite{bib:gronau93}, is maximal for $\delta = 0$.

 In this note, we take a somewhat different point of view and try to gain
 insight into the actual value of the strong phases difference
 $\delta$, without {\it a priori} assumptions on their size, assuming
 that both the mixing-induced and the direct CP violation terms will be
  measured from the time-dependent CP asymmetry of $\bbartopipi$.
 Similarly to most previous analyses, we assume 
the approximation where the {\it top} mediated penguin 
dominates (in which case the electroweak phase of the penguin can be 
identified with the angle $\beta$) and adopt the model dependent estimate of
 the 
ratio $\Apsuat$ provided by factorization. This ratio then depends on both
 the ``true'' value 
of $\alpha$ and on $\beta$.
 Using this relation, we evaluate the correction to the 
``measured'' value of $\alpha$ (obtained from the fit of the 
time-dependent rates in $\bbartopipi$ ), 
required to determine the ``true'' value of $\alpha$, as a function of
$\beta$ alone. 
For this correction we use the exact expression for the penguin contribution 
to the time asymmetry, so that the analysis potentially applies also to the 
case of large penguin-to-tree amplitude ratio. We also gain information 
on the effect of the strong phase $\delta$, corresponding to the different 
possible values of $\beta$.
To summarize, from a time-dependent analysis of $\btopipi$, combined with a 
clean and accurate independent measurement of the angle $\beta$, we show
that one can in principle:
\begin{itemize}
  \item extract the measured uncorrected value of $\alpha$
  \item evaluate (a model dependent) correction $\Delta \alpha$ 
  \item gain insight into the role played by $\delta$ in $\bbartopipi$.
\end{itemize}
 The price is the introduction of a modest, model-dependent, correlation 
between two otherwise independent measurements, and the estimate of the 
ratio $\Apsuat$ from factorization. 

\section{Model-dependent evaluation of $\Apsuat$}
To make the presentation self-contained, this section briefly reviews 
the basic points of the derivation of the penguin-to-tree ratio for 
$\bbartopipi$ in the factorization approach.

The relevant effective weak Hamiltonian can be parameterized as 
\cite{buras3,ciuchini,desh2}: 
\begin{equation} H_W^{eff}=\frac{4G_F}{\sqrt 2}\left[V_{ub}V_{ud}^*
\left(c_1O_1+c_2O_2\right)-V_{tb}V_{td}^*\sum_i c_iO_i\right]
+h.c.,\label{hamilt}\end{equation}
where $c_i$ are short-distance Wilson coefficients defined at a scale $\mu$ 
of the order of the heavy quark mass $m_b$, and $O_i$ are a set of local 
quark operators with the appropriate quantum numbers. The first two terms 
in (\ref{hamilt}) represent the tree diagrams while the other ones are the 
contributions of strong and electroweak penguin diagrams. For simplicity we 
can neglect electroweak penguins, which are found to give a small contribution 
to the mode $\bbartopipi$ of interest here \cite{kramer2,desh2,gronau95}. 
In any case they could be included following the remarks in \cite{aleksan}. 
Retaining only strong penguin operators, the sum in (\ref{hamilt}) reduces to 
$i\leq 6$ and the explicit expressions of the relevant $O_i$ are: 
\begin{equation}\begin{array}{l}
O_1={\bar d}^{\alpha}\gamma_\mu Lu^{\beta}{\bar u}^{\beta}\gamma^{\mu}L
b^{\alpha};\\
O_3={\bar d}\gamma_\mu Lb\sum_{q^\prime}{\bar q}^\prime\gamma^{\mu}Lq^\prime;\\
O_5={\bar d}\gamma_{\mu}Lb\sum_{q^\prime}{\bar q}^\prime\gamma^{\mu}Rq^\prime;
\end{array}
\qquad\qquad\begin{array}{l}
O_2={\bar d}\gamma_{\mu}Lu{\bar u}\gamma^{\mu}Lb;\\
O_4={\bar d}^{\alpha}\gamma_{\mu}Lb^{\beta}\sum_{q^\prime}{\bar q}^{\prime
\beta} 
\gamma^{\mu}Lq^{\prime\alpha};\\
O_6={\bar d}^{\alpha}\gamma_{\mu}Lb^{\beta}\sum_{q^\prime}{\bar q}^{\prime
\beta}\gamma^{\mu}Rq^{\prime\alpha},
\end{array}\label{operators}\end{equation}
where $\alpha,\beta$ are color indices; $L,R=\frac{1}{2}\left(1\mp\gamma_5
\right)$; and $q^\prime$ runs over all quark flavors. For the corresponding 
Wilson coefficients 
at the scale $\mu\sim m_b=4.8\hskip 2pt GeV$ we use the values \cite{desh2}
\begin{equation}\begin{array}{l}
c_1=-0.315;\\ c_4=-0.0373;\end{array}
\qquad\qquad\begin{array}{l}
c_2=1.150;\\ c_5=0.0104;\end{array}
\qquad\qquad\begin{array}{l}
c_3=0.0174;\\ c_6=-0.0459.\end{array}\label{coefficients}\end{equation}

In the factorization hypothesis, after Fierz reordering the operator $O_6$, 
one directly obtains from (\ref{hamilt}) and (\ref{operators})
\begin{equation}\begin{array}{l}
\langle\pi^+\pi^-\vert H_W^{eff}\vert {\bar B}^0\rangle=i\frac{G_F}{\sqrt 2}
\left[\left(m_B^2-m_\pi^2\right)f_\pi F_0^{B\to\pi}(m_\pi^2)\right]
\left\{V_{ub}V_{ud}^*
\left(\frac{1}{N}c_1+c_2\right)\right . \\
\\
\left . 
 -V_{tb}V_{td}^*\left[\left(\frac{1}{N}c_3+c_4\right)+\left(\frac{1}{N}c_5+
c_6\right)\frac{2m_\pi^2}{\left(m_b-m_u\right)\left(m_d+m_u\right)}\right]
\right\},\end{array}\label{factor}\end{equation}
where $N$ is the number of colors. The following definitions have been used:
\begin{equation}
\langle\pi^-\vert{\bar d}\gamma_\mu Lu\vert 0\rangle=-\langle\pi^-\vert 
{\bar d}\gamma_\mu Ru\vert 0\rangle=\frac{i}{2}f_\pi\left(p_{\pi}\right)_\mu
\label{effepi}\end{equation}
with $f_\pi=132\hskip 2pt MeV$ the pion decay constant, so that from quarks 
equations of motion
\begin{equation}\langle\pi^-\vert{\bar d}Lu\vert 0\rangle=-\langle\pi^-\vert
{\bar d}Ru\vert 0\rangle=\frac{i}{2}\frac{m_\pi^2f_\pi}{m_d+m_u};
\end{equation} 
and for the ${\bar B}^0\to\pi^+l^-\nu_l$ matrix elements \cite{bauer}:
\begin{equation} \begin{array}{l}
  \langle\pi^+\vert{\bar u}\gamma_\mu Lb\vert{\bar B}^0\rangle
= \langle\pi^+\vert{\bar u}\gamma_\mu Rb\vert{\bar B}^0\rangle= \\
\\
\frac{1}{2}\left[\left(p_B+p_\pi\right)_\mu+\frac{m_B^2-m_\pi^2}{q^2}
q_\mu\right]F_1^{B\to\pi}(q^2)+\frac{1}{2}\frac{m_B^2-m_\pi^2}{q^2}q_\mu
F_0^{B\to\pi}(q^2),
\end{array}\label{ffactors}\end{equation}
where $q=p_B-p_\pi$, so that from equations of motion
\begin{equation}\langle\pi^+\vert{\bar u}Lb\vert{\bar B}^0\rangle=
\langle\pi^+\vert{\bar u}Rb\vert{\bar B}^0\rangle=\frac{1}{2}
\frac{m_B^2-m_\pi^2}{m_b-m_u}F_0^{B\to\pi}(q^2).\end{equation}
\par 
With both the coefficients $c_i$ and the matrix elements of $O_i$ real, the 
penguin amplitude in Eq.~(\ref{factor}) has the (weak) phase of 
$V_{tb}V_{td}^*$. It is well-known that, by the unitarity of the CKM 
matrix, the $top$ penguin dominance occurs if the difference between the 
$c$- and $u$-mediated penguins can be neglected, which is the case for 
$m_c^2/m_b^2\simeq 0$. However, phenomenological estimates indicate the values 
$m_c=1.3-1.5\hskip 2pt GeV$ and $m_b=4.5-5\hskip 2pt GeV$ \cite{pdg,koide}. 
Thus, the verification of such approximation is clearly a point deserving 
further analysis.\footnote{An attempt of extracting $\sin 2\alpha$ from 
$\bbartopipi$ relaxing this approximation has been recently discussed in 
\cite{mannel}, assuming the approximate linear expansion in $\Apsuat$ of 
the CP asymmetry and the SU(3) relation to the $B\to \pi K$ rate.} As a 
matter of fact, the consistent treatment of both the Wilson coefficients and 
the matrix elements of operators at next-to-leading order in QCD provides 
modifications of order $\alpha_s$ to the simple structure of 
Eqs.~(\ref{coefficients}) and (\ref{factor}).
In particular it introduces imaginary 
parts into the coefficients \cite{kramer2,desh2} which determine final state 
strong interactions phases at the perturbative quark level.
The actual value of such phase is somewhat 
parameter-dependent, although generally quite small, so 
that in the following we will consider the strong interaction phase as a free 
parameter with {\it a priori} arbitrary possible values including (but in 
principle not coinciding with) the perturbative effects.

The general decomposition of the decay amplitude for $\bbartopipi$ in terms of 
tree plus penguin contributions reads: 
\begin{eqnarray}A&=&A(B^0\to\pi^+\pi^-)=A_T \: e^{\: i(\delta_T+\phi_T)}+
A_P \: e^{\: i(\delta_P+\phi_P)}\nonumber\\
{\bar A}&=&A({\bar B}^0\to\pi^+\pi^-)=A_T \: e^{\: i(\delta_T-\phi_T)}+
A_P \: e^{\: i(\delta_P-\phi_P)},\label{amplis}\end{eqnarray}
where $(\phi_T,\phi_P)$ and $(\delta_T,\delta_P)$ are weak and strong 
phases, respectively. In the {\it top dominance} approximation, (\ref{amplis}) 
can be rewritten as
\begin{eqnarray}A&=&\vert V_{ub}^*V_{ud}\vert \: e^{\: i(\delta_T+\gamma)} \:T+
\vert V_{tb}^*V_{td}\vert \: e^{\: i(\delta_P-\beta)} P\nonumber\\
{\bar A}&=&\vert V_{ub}V_{ud}^*\vert \: e^{\: i(\delta_T-\gamma)} \: T +
\vert V_{tb}V_{td}^*\vert \: e^{\: i(\delta_P+\beta)} \: P, 
\label{amplit}\end{eqnarray}
where the CKM entries have been factored from the pure hadronic matrix
elements $T$ and $P$. In the leading order Wolfenstein parameterization, the 
unitarity triangle angles :
\begin{equation} \alpha=arg\left(-\frac{V_{td}V_{tb}^*}{V_{ud}V_{ub}^*}
\right);\qquad\quad \beta=arg\left(-\frac{V_{cd}V_{cb}^*}{V_{td}V_{tb}^*}
\right);\qquad\quad \gamma=arg\left(-\frac{V_{ud}V_{ub}^*}{V_{cd}V_{cb}^*}
\right)\label{angles}\end{equation}
are related to the parameters $\rho$ and $\eta$ as
\begin{equation}
\tan\alpha=\frac{\eta}{\eta^2-\rho\left(1-\rho\right)};\qquad 
\tan\beta=\frac{\eta}{1-\rho};\qquad \tan\gamma=\frac{\eta}{\rho}.
\label{roeta}\end{equation}

Comparing (\ref{amplit}) to (\ref{factor}) one directly reads the expressions 
of $T$ and $P$ in the factorization approach:
\begin{eqnarray}
 T&=&\:\:\:\:\:\: \frac{G_F}{\sqrt 2}\left[\left(m_B^2-m_\pi^2\right)
f_\pi F_0^{B\to\pi}(m_\pi^2)\right]\left(\frac{1}{N}c_1+c_2\right),
\label{t}\end{eqnarray}
\begin{eqnarray}
\:\:\:\:\:\:\: P&=&-\frac{G_F}{\sqrt 2}\left[\left(m_B^2-m_\pi^2\right)
f_\pi F_0^{B\to\pi}(m_\pi^2)\right]\nonumber \\
&\times&\left[\left(\frac{1}{N}c_3+c_4\right)+\left(\frac{1}{N}c_5+c_6\right)
\frac{2m_\pi^2}{\left(m_b-m_u\right)\left(m_d+m_u\right)}\right].
\label{p}\end{eqnarray}
From (\ref{t}) and (\ref{p}) one can notice the property of the ratio $P/T$ 
of being independent of particular models for the $B\to\pi$ form factors. 
Thus, apart from the factorization hypothesis itself, it only depends on 
the short-distance coefficients which can slightly vary according to the 
choice of the scale $\mu$, and on the current quark masses in the last term 
of (\ref{p}). We choose $m_b=4.8\hskip 2pt GeV$, $m_d+m_u=15\hskip 2pt MeV$ 
so that, with the coefficients (\ref{coefficients}) and $N=3$ (the dependence 
on $N$ is weak), one finds, as in \cite{bib:kramer1,aleksan},
$P/T\simeq 0.055$ .
Reasonable variations of quark masses around the chosen 
values, taking into account the spread of the various determinations, can 
affect the term mentioned above by about 30\%. However, being anyway 
suppressed in (\ref{p}) by the large $m_b$ and small short distance 
penguin coefficients, 
this reflects into an uncertainty on the ratio $P/T$ of the order of 10\%.

\section{Discussion on the penguin correction $\Delta \alpha$}

\begin{figure}[htbp]
  \begin{center}
      \mbox{\epsfig{figure=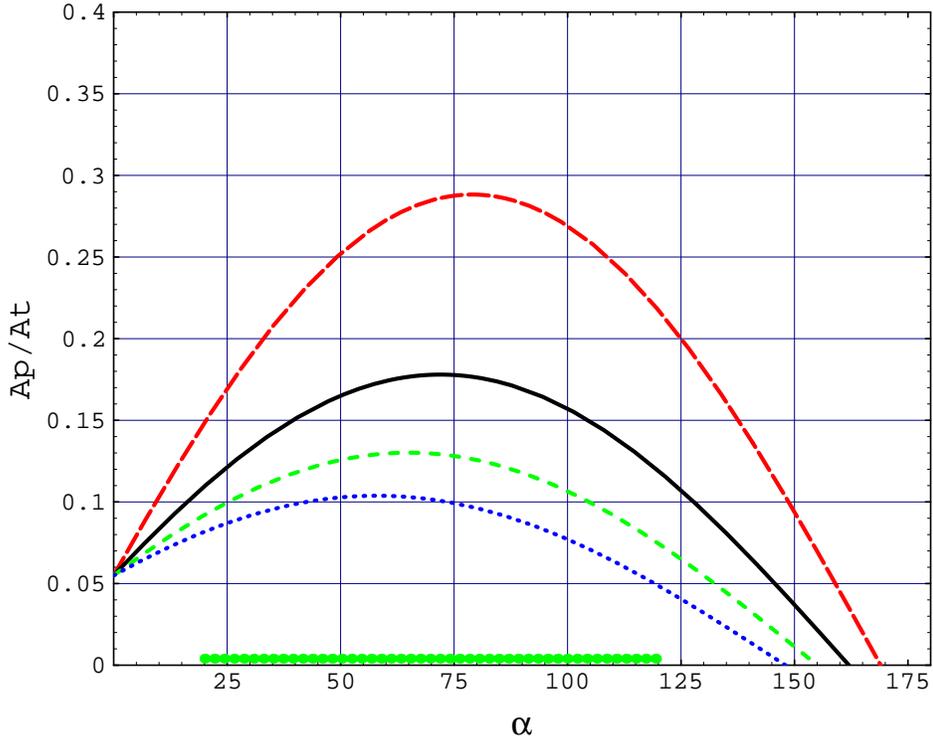,%
         height=5.5in,bbllx=83pt,bblly=150pt,bburx=576pt,bbury=653pt}}
  \end{center}
  \caption[  $\Apsuat$ plotted against the weak phase $\alpha$ 
 for four different values of $\beta = 11^{\circ}$ (dashed upper curve),
 $18^{\circ}$ (solid), $25^{\circ}$ (dashed), $32^{\circ} (dotted) \: \: \:$.
 ]{\label{fig:apt}
 $\Apsuat$ plotted against the weak phase $\alpha$ 
 for four different values of $\beta = 11^{\circ}$ (dashed upper curve),
 $18^{\circ}$ (solid), $25^{\circ}$ (dashed), $32^{\circ} (dotted) \: \: \:$.
           }
\end{figure}

\begin{figure}[htbp]
  \begin{center}
      \mbox{\epsfig{figure=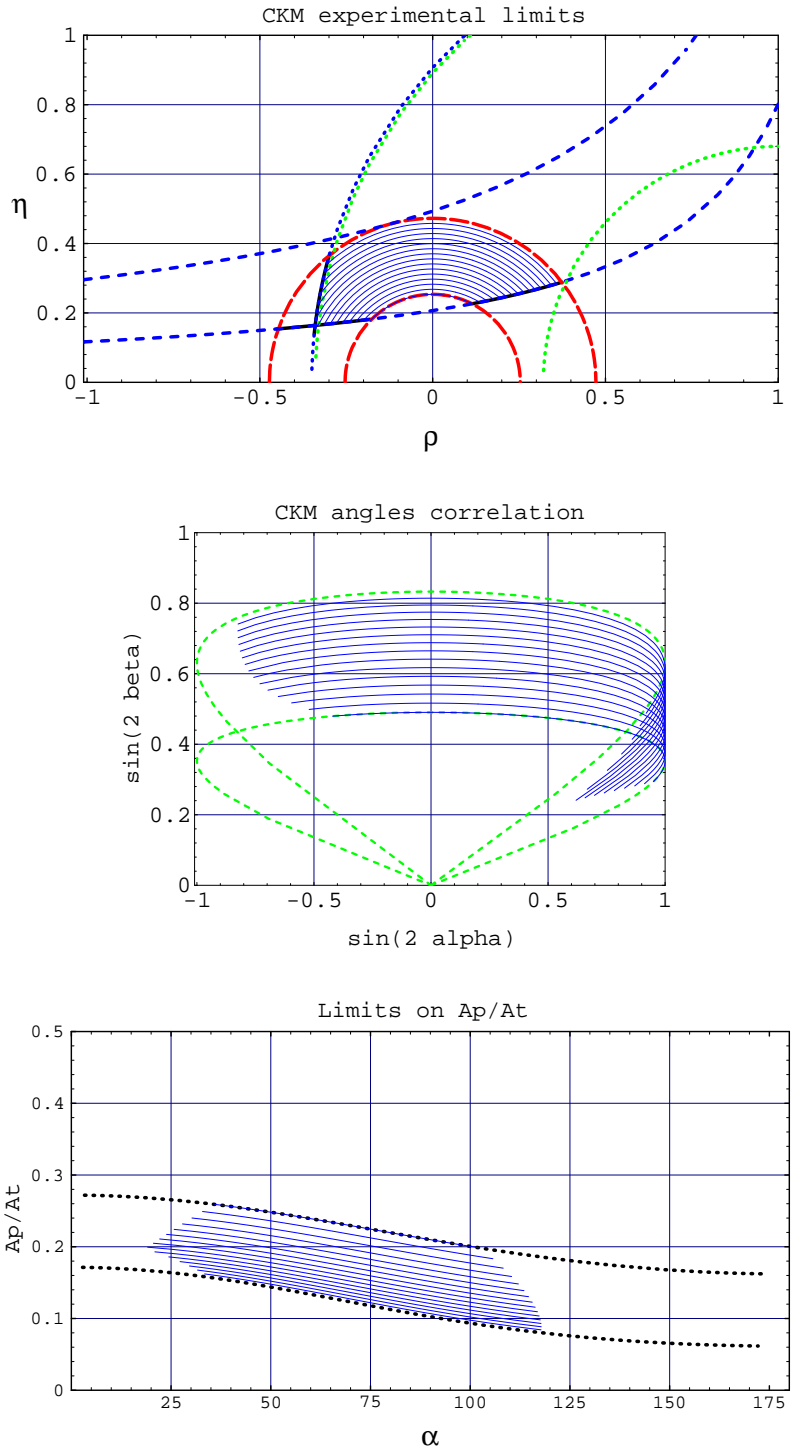,%
           height=7.5in,bbllx=83pt,bblly=70pt,bburx=576pt,bbury=683pt}}
  \end{center}
  \caption[ 
  (a) current experimental limits in the ($\rho,\eta$) plane
; (b) correlation between $\sen2beta$ and $\sin 2 \alpha \:$ 
; (c) current bounds on $\Apsuat$ plotted vs. $\:\alpha$ for $P/T = 0.055 \:$.
]{\label{fig:bounds}
  (a) current experimental limits in the ($\rho,\eta$) plane
; (b) correlation between $\sen2beta$ and $\sin 2 \alpha \:$ 
; (c) current bounds on $\Apsuat$ plotted vs. $\:\alpha$ for $P/T = 0.055 \:$.
           }
\end{figure}

\begin{figure}[htbp]
  \begin{center}
      \mbox{\epsfig{figure=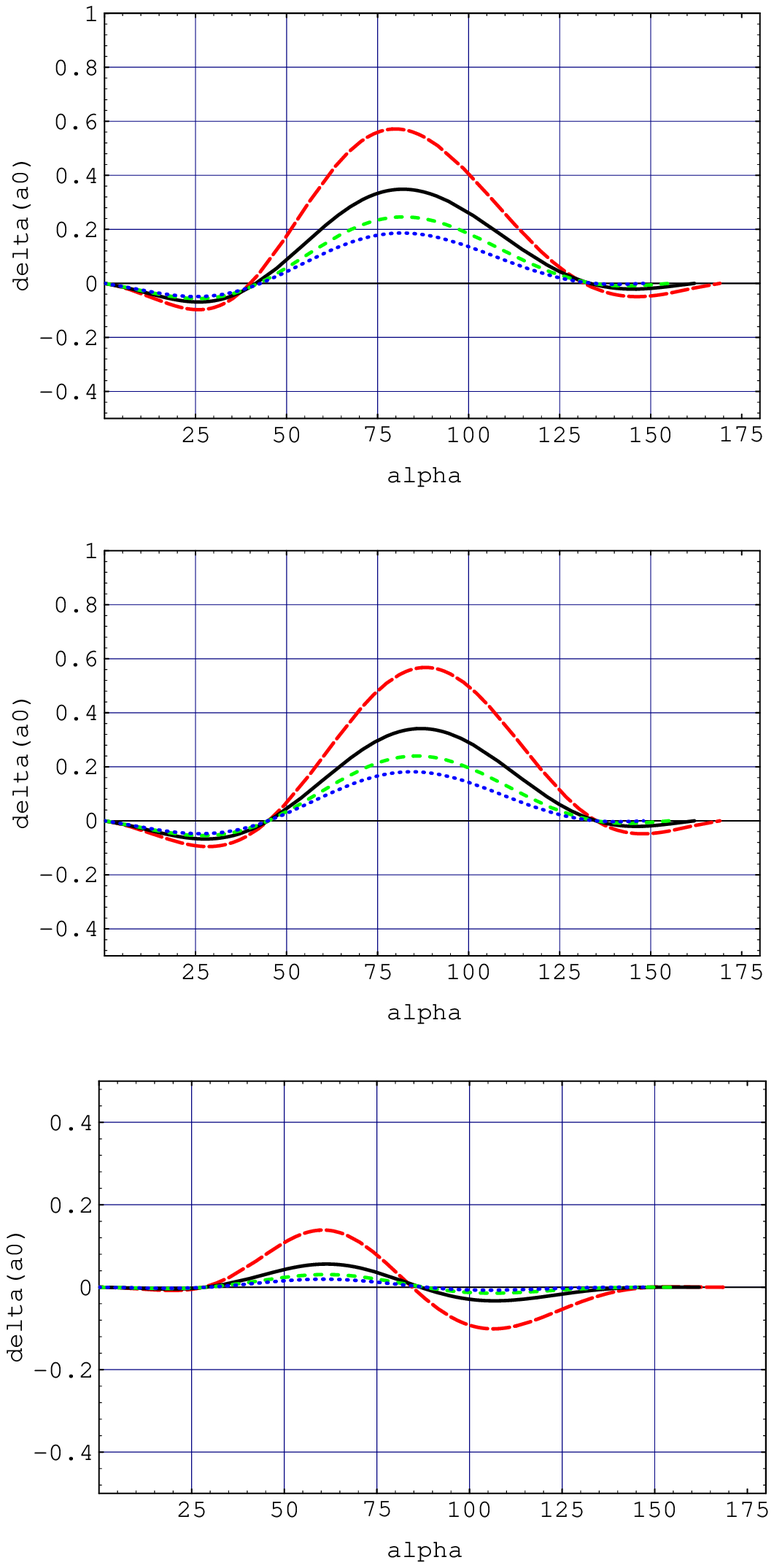,%
           height=6.4in,bbllx=103pt,bblly=100pt,bburx=596pt,bbury=603pt}}
  \end{center}
  \caption[ 
 $\Delta \azero$ correction plotted against the weak phase $\alpha$ for $\delta = 0$ and 
 four different values of $\beta = 11^{\circ}$ (dashed upper curve),
 $18^{\circ}$ (solid), $25^{\circ}$ (dashed), $32^{\circ} (dotted) \: \: \:$
 (a) : exact correction ;$\: \: \: $
 (b) : approximate correction to first order in $\Apsuat \:$; 
 (c) : difference (a)-(b). 
]{\label{fig:1st2nd}
 $\Delta \azero$ correction plotted against the weak phase $\alpha$ for $\delta = 0$ and 
 four different values of $\beta = 11^{\circ}$ (dashed upper curve),
 $18^{\circ}$ (solid), $25^{\circ}$ (dashed), $32^{\circ} (dotted) \: \: \:$
 (a) : exact correction ;$\: \: \: $
 (b) : approximate correction to first order in $\Apsuat \:$; 
 (c) : difference (a)-(b). 
           }
\end{figure}

 The time-dependent decay rates are given by :
\begin{equation}
  \Gamma( B^{0}(\overline {B^{0}}) \to {\pi^{+}} {\pi^{-}} ) 
= \frac{\mid \overline{A} \mid ^{2} + \mid A \mid ^{2}}{2} 
\: e^{-\Gamma \mid t \mid} \: ( 1 \pm \azero \sin{\Delta mt}
                                     \pm \bzero cos{\Delta mt} )
  \label{eq:time_rate}
\end{equation}
where the two real coefficients $\azero$ and $\bzero$ are :\\
\begin{eqnarray}
  \label{eq:a0b0}
     &  &  \mbox{}
     \azero = \frac{2 \: {\cal I}m \: (e^{2 i \beta} A \overline{A}^{*}) }
                   {\mid A \mid ^{2} + \mid \overline{A} \mid ^{2}} 
     \nonumber \\
     &  &  \mbox{}
     \\
     &  &  \mbox{}
     \bzero = \frac{\mid A \mid ^{2} - \mid \overline{A} \mid ^{2}}
                   {\mid A \mid ^{2} + \mid \overline{A} \mid ^{2}} 
     \nonumber
\end{eqnarray}
 In terms of the ratio of the penguin to tree contribution 
 $\Apsuat$, the two coefficients $\azero$ and $\bzero$ 
can be written as :\\
\begin{eqnarray}
  \label{eq:a0dev}
     \azero = \frac{ -\sin 2 \alpha \, + 2 \:(\Apsuat) \cos \delta \: \sin \alpha }
                   { 1 - 2 \:(\Apsuat) \: \cos \delta \: \cos \alpha + (\Apsuat)^{2} }
\end{eqnarray}
\begin{eqnarray}
  \label{eq:b0dev}
     \bzero = \frac{ - 2 \:(\Apsuat) \: \sin \delta \: \sin \alpha }
                   { 1 - 2 \:(\Apsuat) \: \cos \delta \: \cos \alpha + (\Apsuat)^{2} }
\end{eqnarray}
 \\
 where $\delta \equiv \deltat - \deltap$ 
 is the strong phases difference and, as in eq.(\ref{amplit}), 
 $\: \phit - \phip = \gamma + \beta = \pi - \alpha $ in the SM.
 When $\Apsuat = 0 \:$, the $\bzero$ term vanishes and 
 $\alpha$ is extracted from the simple relation 
 $\azero = - \sin 2 \alpha \,$. Else, a $\Delta \alpha $ correction term 
 which depends both on $\alpha$ and $\cos \delta$ , has to be added.
 The $\Delta \alpha $ correction is therefore maximal 
 for $\delta = 0$, as pointed out previously.\\
 The ratio of the CKM matrix elements in $\Apsuat$ can be expressed
in terms  of the angles $\alpha$ and $\beta$ of the unitary triangle 
as :
\begin{equation}
  \frac{\mid \Vtb \Vtdstar \mid}{\mid \Vub \Vudstar \mid}  =
  \frac{\sin \gamma}{\sin \beta} \: 
  \label{eq:seni}
\end{equation}
and therefore, in terms of the ratio of pure hadronic matrix elements $P/T$
\begin{equation}
  \frac{\Ap}{\At} = ( \sin \alpha \cot \beta + \cos \alpha ) \: (\frac{P}{T})
  \label{eq:Apteq}
\end{equation}
 The expected value of the ratio $\Apsuat$ --- calculated using 
 the  $P/T$ value obtained in the previous section ---
 is plotted in fig.~\ref{fig:apt} as a function of $\alpha$ 
 for four different values  of $\beta$. 
 The curves in fig.~\ref{fig:apt} are drawn for   
 $\beta = 11^{\circ}, 18^{\circ}, 25^{\circ}, 32^{\circ}$, respectively  
 and the highest (lowest) curve lies close 
 to the present experimental lower (upper) limit on $\beta$
 \cite{bib:Ali96},\cite{bib:AliDec96},\cite{bib:Rosner96}.
 In the same picture, the current experimental bounds on $\alpha$  are
 superimposed onto the horizontal axis. \\
 From inspection of fig.~\ref{fig:apt}, we would conclude
 that the value of $\Apsuat$ predicted by the model could be as large as
 0.3 if the value of  $\beta$ 
 would turn out to be close to its present lower limit.
 However, we must keep in mind that the allowed ranges for $\sin 2 \alpha$
 and $\sen2beta$ are correlated through the common dependence on $\rho$ and
 $\eta$ (see figs.~\ref{fig:bounds}(a),(b)).
 Therefore, the values of 
 $\alpha$ and $\beta$ are allowed to cover a {\it subset} of the 
 region bounded in fig.~\ref{fig:apt} by the envelope of the
 $\beta$ curves and by the $\alpha$ limits on the ascissa.
 This is best shown in fig.~\ref{fig:bounds}(c) where 
 the present limits on the CKM triangle
 \cite{bib:Ali96},\cite{bib:AliDec96},\cite{bib:Rosner96}
 are used to bound the shaded area which represents the allowed
 value of $\Apsuat$ as a function of $\alpha$. From the picture, we see that 
 the penguin to tree amplitude ratio can, at present, vary from a minimum of 
 0.08 to a maximum of 0.26, approximately.\\
 From equation (\ref{eq:a0dev}),
 the correction $ \Delta \azero$ 
 to the observed value of the $\azero$ coefficient
\begin{equation}
  \label{eq:deltazero}
\Delta \azero = \azero + \sin 2\alpha
   \nonumber 
\end{equation}
is plotted in fig.~\ref{fig:1st2nd}(a) for $\delta = 0$, as a
function of $\alpha$.
For the same value of $\delta$, fig.~\ref{fig:1st2nd}(b) shows
 the  $ \Delta \azero$
correction obtained, as in \cite{bib:gronau93}, 
by expanding equation (\ref{eq:a0dev})
 to first order in $\Apsuat$
\begin{equation}
  \label{eq:firstorder}
\Delta \azero = - 2 \:(\Apsuat) \: \sin \alpha \: \cos 2\alpha \: \cos \delta
\end{equation}
 and taking $\delta = 0$, while fig.~\ref{fig:1st2nd}(c) 
 shows the difference between the exact and the first order result.
  A common feature is that the  $\Delta \azero$ correction is positive in the 
 interval $ \pi/4 < \alpha < 3\pi/4 $  while it has the opposite sign and 
 is smaller in magnitude, outside this range.   
 However, while the first order expansion predicts a correction
 which vanishes for $\alpha = \pi/4$ and $\alpha = 3\pi/4$ ( for any value
 of $\beta$ ) and is maximal for $\alpha = \pi/2$, the second order term
 introduces a dependence on $\beta$ in the position of the zeroes and 
 of the maximum of the  $ \Delta \azero$ correction.\
 Within the presently allowed range of
 $\alpha$, the largest difference in  $ |\Delta \azero| $ between the 
 exact $ \Delta \azero$ correction from eq.(\ref{eq:a0dev}) and the 
 first order approximation of eq.(\ref{eq:firstorder})
 shows up ( see fig.~\ref{fig:1st2nd}(c)) at 
 $\alpha \approx 60^{\circ}$ and 
 $\alpha \approx 105^{\circ}$ approximately, while it is negligible 
 for $\alpha$ values close to $\pi/2$.
 It is worth noticing that we have used a {\it positive} value for $P/T$ as 
 provided by the model. When reversing its sign, 
the correction $\Delta \azero$ changes sign too.\\
 Rewriting the (second order) expressions for the two parameters 
$\azero$ and $\bzero$
 in terms of $\alpha $ ,$\: \delta \:$ and $\Apsuat \:$
 in eqs.(\ref{eq:a0dev}) and (\ref{eq:b0dev}) as :
\begin{eqnarray}
  \label{eq:sys1}
  \azero + \sin 2 \alpha -  2 \: (\Apsuat) \: \cos \delta \:
       ( \: \azero \cos \alpha \: +  \sin \alpha \:) + \azero \: (\Apsuat)^{2}
  = \: 0
 \\
  \label{eq:sys2}
  \bzero -  2 \: (\Apsuat) \: ( \: \bzero \: \cos \delta \: \cos \alpha \:
  - \sin \delta \: \sin \alpha \:) + \bzero \: (\Apsuat)^{2} = \: 0
\end{eqnarray}
 and inserting $\Apsuat$ from (\ref{eq:Apteq}) into eqs.(\ref{eq:sys1}),
 (\ref{eq:sys2}) with the value of $P/T$ provided by the model, 
 we get {\it two equations} in the {\it two unknowns}
 $\alpha$ and $\delta$, both
 a function of the $\beta$ angle :  
 $\alpha = \alpha(\beta)$ and $\delta = \delta(\beta)$ . \\

 Once $\beta$ is known from an independent measurement
 ( e.g.: from the decay $\btojpsi$ ) 
 and the two parameters $\azero$ and $\bzero$ are fitted from the 
 $\bbartopipi$  time-dependent decay rates, we can in principle extract 
 the correct value of $\alpha$, from the data, {\it together with}
 a measurement of the relative strong phase $\delta$.

\section{Numerical solutions for $\Delta \alpha $ and $\delta$}
\label{sec:numerical}

 For a given value of $\beta$, we solve {\it numerically} 
 the system of two equations (\ref{eq:sys1}) and (\ref{eq:sys2})  
 in the two unknowns $\alpha$ and $\delta$ and find the
 correction  $\Delta \alpha $ to be applied to the {\it measured} value
 $\alpham$ of $\alpha$  defined as : 
\begin{equation}
  \label{eq:alpham}
\Delta \alpha = \alpham - \alpha
   \nonumber 
\end{equation}
where $\alpham$ is inferred from the 
fitted value of the $\azero$ coefficient, with no penguin corrections, via the
relation :
\begin{equation}
  \label{eq:amdef}
 - \sin 2 \alpham = \azero
   \nonumber 
\end{equation}
 Since the value of the measured asymmetry parameter 
 $\azero$ is related to the weak phase $\alpham$ via the above
 circular relation, one cannot distinguish between the case where $\alpham$ 
 falls into the $[\:45^{\circ},135^{\circ}\:]$ interval or outside.
 Therefore, for  $ -1 < \azero < 1 \: \: \: $, two numerical solutions
 for the true value of $\alpha$ are found.
 The first one corresponds to the above interval for $\alpham$,
 while the second solution
 belongs to the $ [\:0^{\circ},45^{\circ}]$ interval for $\azero < 0 $ 
 and to the $ [\:135^{\circ},180^{\circ}\:] \:$ interval for $\azero > 0 $ .\\

 Before taking into account the constraints from the present experimental
 bounds on $\alpha$ and $\beta$,
 we first examine the predictions for both solutions i.e.:  with $\alpha$ 
 spanning the full $[0 \: ,\: \pi - \beta ]$ interval.\\

\begin{figure}[htbp]
  \begin{center}
      \mbox{\epsfig{figure=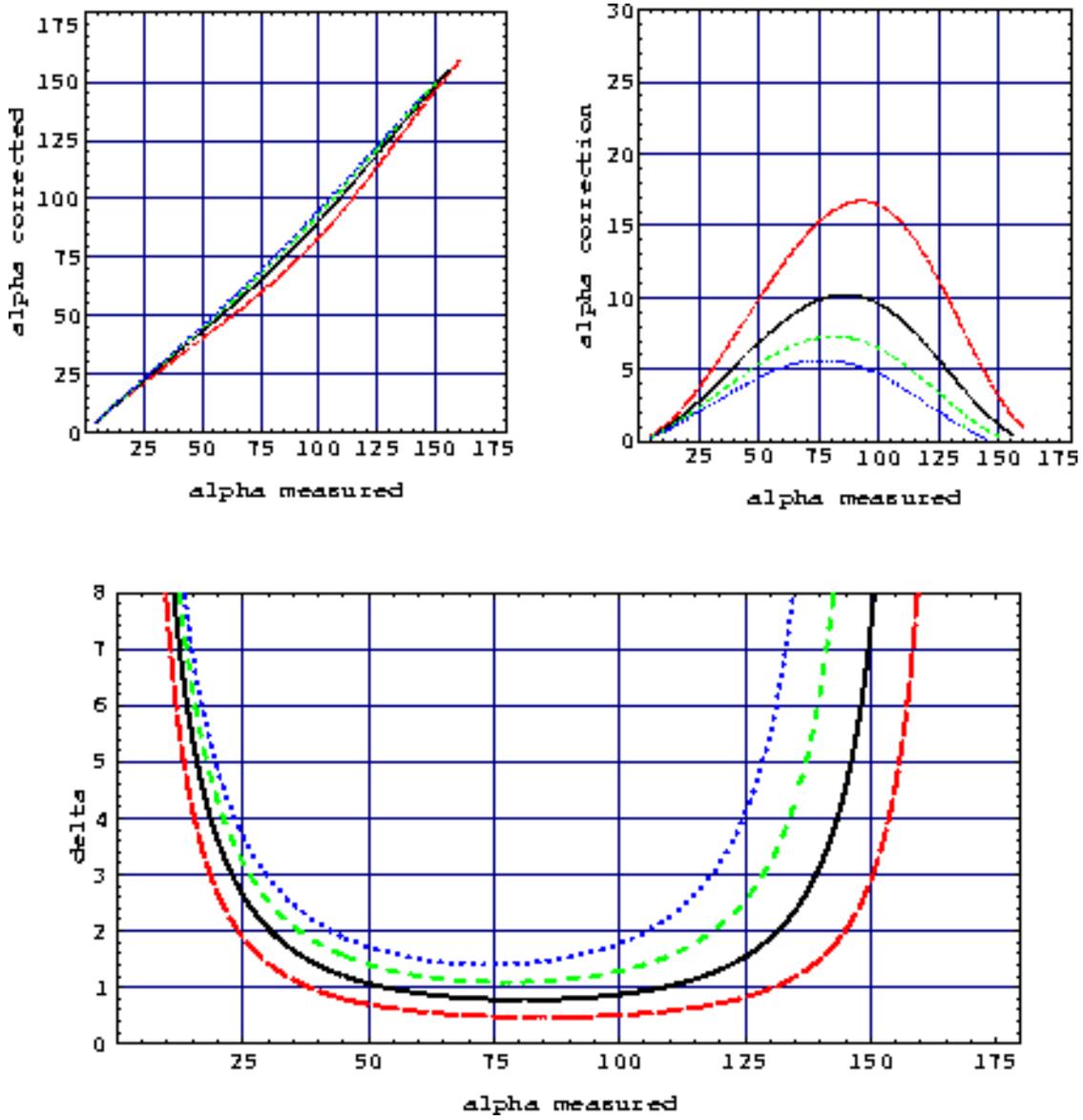,%
           height=7.0in,bbllx=103pt,bblly=120pt,bburx=596pt,bbury=623pt}}
  \end{center}
  \caption[
 For four different values of $\beta = 11^{\circ}$ (dashed),
 $18^{\circ}$ (solid), $25^{\circ}$ (dashed), $32^{\circ}$ (dotted) 
 and $ \: P/T = 0.055 \: \:$ :
 (a) ``$\alpha$ corrected'' vs.  ``$\alpha$ measured'' ;
 (b) $\Delta \alpha$ correction vs.  ``$\alpha$ measured'';
 (c) relative strong phase $\delta$ vs. ``$\alpha$ measured'' ( see
 section 4 for a detailed explanation of this picture ).
  ]{\label{fig:twoinone}
 For four different values of $\beta = 11^{\circ}$ (dashed),
 $18^{\circ}$ (solid), $25^{\circ}$ (dashed), $32^{\circ}$ (dotted) 
 and $ \: P/T = 0.055 \: \:$ :
 (a) ``$\alpha$ corrected'' vs.  ``$\alpha$ measured'' ;
 (b) $\Delta \alpha$ correction vs.  ``$\alpha$ measured'';
 (c) relative strong phase $\delta$ vs. ``$\alpha$ measured'' ( see
 section 4 for a detailed explanation of this picture ).
           }
\end{figure}

\begin{flushleft} \small \bf
 The penguin correction $\Delta \alpha$ 

\end{flushleft}
The {\it ``penguin-corrected''} weak phase $\alpha$ 
 and the correction  $\Delta \alpha $ resulting from our analysis
 are plotted in  fig.~\ref{fig:twoinone}(a) and fig.~\ref{fig:twoinone}(b),
 respectively,  against the
 {\it measured ( uncorrected )} value $\alpham$ in $\btopipi$ 
 for four different values 
 of $\beta = 11^{\circ}, 18^{\circ}, 25^{\circ}, 32^{\circ}$ and 
 with a fixed value of $\bzero = -5.0 \: 10^{-3} $. 
 This value has been chosen as an example 
 where the value of the 
 $\bzero$ parameter is very small, but 
 not identically zero, to exemplify a case where $\delta \neq 0 $.\\

\begin{figure}[htbp]
  \begin{center}
      \mbox{\epsfig{figure=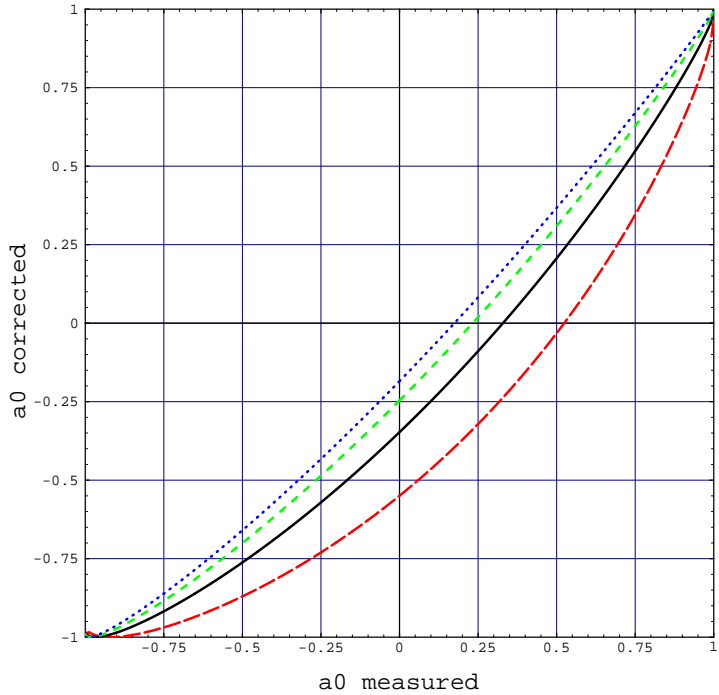,%
        height=4.0in,bbllx=83pt,bblly=180pt,bburx=576pt,bbury=683pt}}
  \end{center}
  \caption[ $-\sin 2 \alpha$ vs. measured $\azero$ 
 for four different values of $\beta = 11^{\circ}$ (dashed),
 $18^{\circ}$ (solid), $25^{\circ}$ (dashed), $32^{\circ} (dotted)$
 with $\bzero = 0$ and $ \: P/T = 0.055 \: \: \:$.
  ]{\label{fig:sin}
            $-\sin 2 \alpha$ vs. measured $\azero$ 
 for four different values of $\beta = 11^{\circ}$ (dashed),
 $18^{\circ}$ (solid), $25^{\circ}$ (dashed), $32^{\circ} (dotted)$
 with $\bzero = 0$ and $ \: P/T = 0.055 \: \: \:$.
           }
\end{figure}

 From inspection of fig.~\ref{fig:twoinone}(a) and fig.~\ref{fig:twoinone}(b)
 we note that :
\begin{itemize}
 \item  the $\Delta \alpha $ correction increases when $\beta$ decreases.
   A maximum  $\Delta \alpha $ correction of about $16.5 ^{\circ}$ 
   is found for $\beta = 11^{\circ}$.
   The correction does not exceed $10^{\circ}$ for $\beta$ values 
   greater than $18^{\circ}$, approximately. 

 \item  maximal $\Delta \alpha $ corrections for different $\beta$ 
   values occur when $\alpham$ lies  within the approximate interval 
   ($ [\:70^{\circ},100^{\circ}]$) and  
   the true value of $\alpha$ corresponding to the maximum of the correction
   is found to increase when $\beta$ decreases. However, 
   to establish the maximal correction -- allowed within the present 
   limits on $\alpha$ and $\beta$ -- one should 
   take into account the correlation between the allowed  
   ranges of the two weak phases, as already pointed out for 
   fig.~\ref{fig:apt}.
\end{itemize}
 The ``penguin-corrected'' value of the asymmetry 
 $-\sin2\alpha$ in $\btopipi$, 
 corresponding to $\alpham$ in the
 $[\:45^{\circ},135^{\circ}\:]$ interval, 
 is plotted in  fig.~\ref{fig:sin}
 against the {\it measured} value of  $\azero$
 for four different values 
 of $\beta = 11^{\circ}, 18^{\circ}, 25^{\circ}, 32^{\circ}$ 
 and with $\bzero = 0$.
 As expected from fig.~\ref{fig:1st2nd}(a) , we note that
 the measured value of $\azero$ is significantly higher than the ``true'' value 
 $- \sin 2 \alpha$ and that the
 correction increases for decreasing $\beta$ values.
 On the contrary, when $\alpham$ lies {\it outside } 
 the above interval, the correction ( not shown in this picture) to 
 the true value of $\alpha$ is marginal and
 of opposite sign with respect to the previous case
 ( i.e.: the measured value of $\azero$ is slightly larger than 
 the ``true'' value ).\\

\begin{flushleft} \small \bf
 The strong phases difference $\delta$ 
\end{flushleft}

 In the above example, the values of the solution for the strong 
 phases difference $\delta$ corresponding to
 a measured value $\alpham$ are plotted as
 the family of curves in fig.~\ref{fig:twoinone}(c)
 for the same values of $\beta$ used in the previous plots.
 In the picture, we note that 
 the relative strong phase $\delta$ shows a
 {\it broad minimum } which covers most of the region 
 where $\alpha$ is bounded at present.
 In our example, the value of the $\bzero$ parameter being very close to zero, 
 the minimum value of $\delta$ turns out to be quite small ($< 2^{\circ} \:$,
 approximately).
   On the contrary, 
   the value of $\delta$ shows a fast increase when $\alpha$ 
   approaches the two ``geometrical'' limits 0 or  $(\pi - \beta)$. 
   We want to stress that the dependence of $\delta$ on $\alpham$
   in fig.~\ref{fig:twoinone}(c) has no direct physical meaning. Instead, it
   describes the parametric behaviour of the solutions 
   of (\ref{eq:sys1}) and (\ref{eq:sys2})
   for the unknowns $\delta$ and $\alpha$ when the value of the 
   $\bzero$ parameter is 
   {\it kept fixed at a non zero value} and $\azero$ is varied.
   The behaviour of $\delta$ on the ascissa in fig.~\ref{fig:twoinone}(c) 
   can be understood by considering the following (exact)
   relation :
\begin{equation}
  \tan\delta = \frac{\bzero [\cos2\alpha -(\Apsuat)^2]}
                    {\azero [1 + (\Apsuat)^2] + \sin2\alpha}
  \label{eq:tandelta}
\end{equation}
   which can be easily derived from eqs.(\ref{eq:a0dev}) and 
   (\ref{eq:b0dev}). \\
   The ratio $\Apsuat$ is found, from eq.(\ref{eq:Apteq}), to
   vanish in the limit $\alpha \rightarrow (\pi - \beta)$, while 
   $\Apsuat = P/T = 0.055$ for $\alpha = 0 \: \: $ (see also fig.~\ref{fig:apt}).
   Correspondingly, $\: \Delta \azero \rightarrow 0$ 
   for both $\alpha \rightarrow 0$ and  $\alpha \rightarrow (\pi - \beta)$.
   Therefore, when $\alpha$ is close to zero, the numerator of 
   eq.(\ref{eq:tandelta}) is
   almost constant, while the denominator approaches zero since, in this
   case, the $\Delta \azero$ correction is almost negligible and 
   the measured asymmetry $\azero$ does not differ significantly from
   the true asymmetry which is identically zero 
 (see fig.~\ref{fig:twoinone}(b)).\\
   A similar, but not identical behaviour, takes place for 
  $\alpha \rightarrow (\pi - \beta$).
 This accounts for the fast increase of $\delta$ on the final 
 portion of the curves in fig.~\ref{fig:twoinone}(c), 
 drawn for the four different values of $\beta$ 
 when $\alpha$ 
 approaches the respective values  $\pi - \beta$.\\

\begin{figure}[htbp]
  \begin{center}
      \mbox{\epsfig{figure=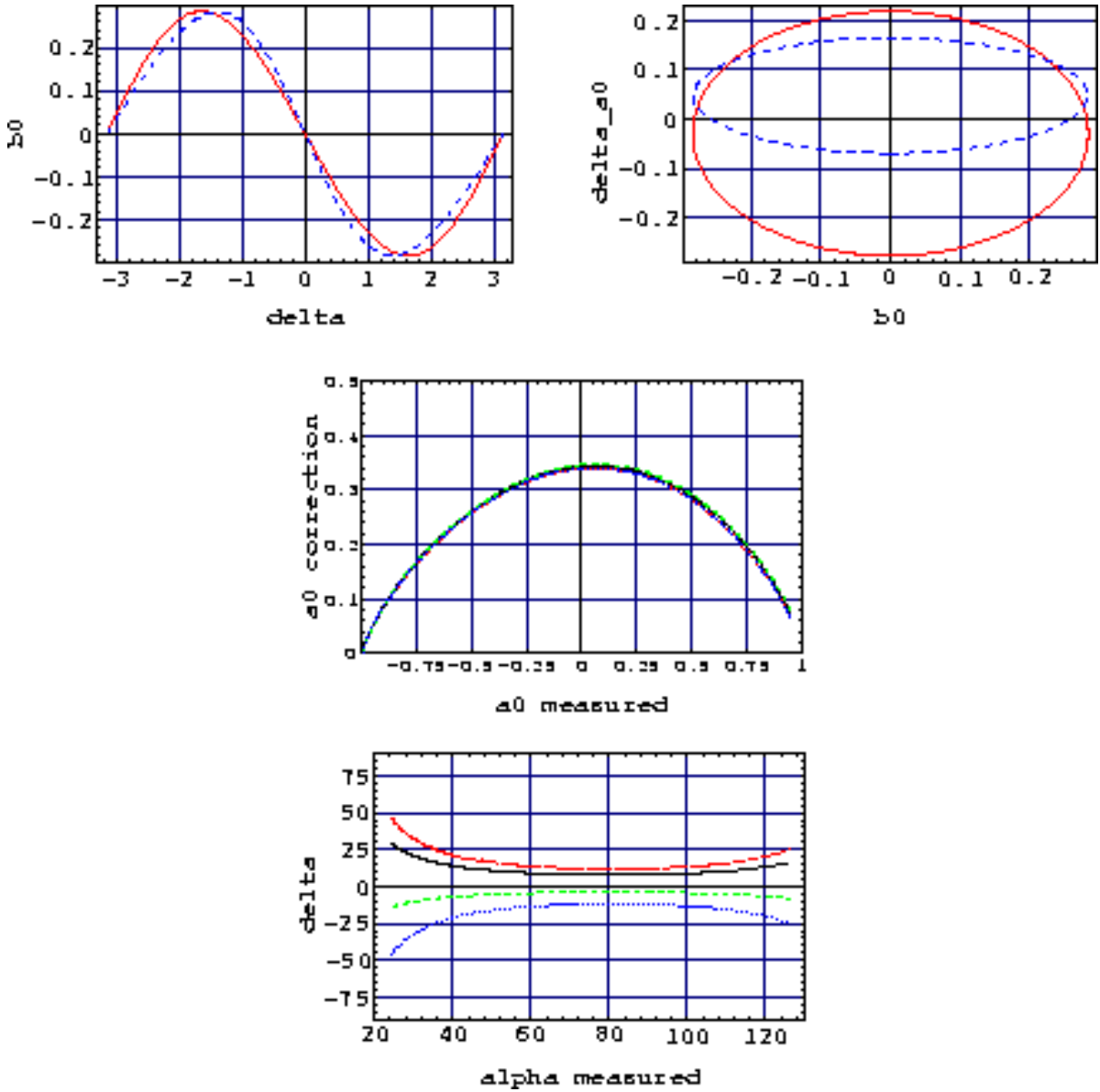,%
           height=7.5in,bbllx=123pt,bblly=150pt,bburx=616pt,bbury=653pt}}
  \end{center}
  \caption[ 
 With $\beta = 18^{\circ}$ in the four pictures :
 $ \: \: $
 (a) $\: \bzero$ vs. $\delta\: \:$ and
 (b) $\: \Delta \azero$ vs. $\bzero\: \:$ 
 for $\alpha = 104.8^{\circ}$ (solid) and $\alpha = 56.5^{\circ}$(dashed).
 $ \: $
 Taking four different values of $\bzero=-0.075$ (dashed),
 $-0.050$ (solid), $+0.025 $ (dashed), $+0.075$ (dotted) : $ \: \: $
 (c) correction $\Delta \azero$ vs. the measured asymmetry $\azero \:$;
 (d) relative strong phase $\delta$ vs. ``$\alpha$ measured''.
 $ \: \: $
]{\label{fig:bsin}
 With $\beta = 18^{\circ}$ in the four pictures :
 $ \: \: $
 (a) $\: \bzero$ vs. $\delta\: \:$ and
 (b) $\: \Delta \azero$ vs. $\bzero\: \:$ 
 for $\alpha = 104.8^{\circ}$ (solid) and $\alpha = 56.5^{\circ}$(dashed).
 $ \: $
 Taking four different values of $\bzero=-0.075$ (dashed),
 $-0.050$ (solid), $+0.025 $ (dashed), $+0.075$ (dotted) : $ \: \: $
 (c) correction $\Delta \azero$ vs. the measured asymmetry $\azero \:$;
 (d) relative strong phase $\delta$ vs. ``$\alpha$ measured''.
 $ \: \: $
           }
\end{figure}

\subsection{ The role of the \bzero parameter }
\label{sec:deltarole}

 The relation between $\bzero$ and $\delta$, for given values
 of $\alpha$ and $\beta$, is given in eq.(\ref{eq:b0dev}). 
 Using the value of $P/T$ from section 2, 
 the numerator of (\ref{eq:b0dev}) is found to dominate
 over the weak $\cos \delta$ dependence of the denominator
 and therefore $\bzero$ turns out to be approximately proportional 
 to $-\sin \delta \:$. This dependence is shown in 
 fig.~\ref{fig:bsin}(a) where, taking $\beta = 18^{\circ} $ 
 and letting $-\pi \le \delta \le \pi$, we plot $\bzero$ as a function of 
 $\delta$ for two values of
 $\alpha = 104.8^{\circ}$ (solid line) 
 and $\alpha = 56.5^{\circ}$(dashed).
 In the same figure, the values of $\delta$ corresponding to
 a minimum or a maximum for $\bzero$ are not too far
 from $\pm \pi/2$ for both curves in our example, as
 the absolute value of the $\bzero$ parameter is maximal for 
 $\cos\delta = 2 (A_{P}/A_{T}) \cos\alpha \: / \:
    [1 + (A_{P}/A_{T})^2] $. \\
 In order to show the dependence of
 the ``penguin-correction'' $\Delta \azero$, defined  
 in eq.(\ref{eq:deltazero}),
 on the actual value of the $\bzero$ parameter, we stick to the 
 above example and plot  
 $\Delta \azero$ as a function of $\bzero$ in fig.~\ref{fig:bsin}(b). 
 For $-\pi \le \delta \le \pi$ , we found that 
 $\Delta \azero$ and $\bzero$ satisfy an elliptic constraint (whose 
 geometrical parameters depend on the choice of $\alpha$ and $\beta$).
  This is readily verified
 using the first order expansion 
  (\ref{eq:firstorder}) in $\Apsuat$ 
 of equation (\ref{eq:a0dev}) and the analogous expansion of
 (\ref{eq:b0dev}). In this approximation, the constraint is simply :
\begin{equation}
 [\frac{\Delta \azero}{2 \: (\Apsuat) \sin\alpha \: \cos 2 \alpha}]^2 + 
 [\frac{\bzero}{2 \: (\Apsuat) \sin\alpha}]^2 = 1 
  \label{eq:ellisse}
\end{equation}
 Using instead the (exact) equations (\ref{eq:a0dev}) and (\ref{eq:b0dev})
 , the constraint is no longer represented by the approximate 
 implicit form (\ref{eq:ellisse}) 
 and the ellipse is shifted along the $\Delta \azero$
 axis as in fig.~\ref{fig:bsin}(b) by a 
 quantity which depends both on $\alpha$ and $\beta$ (see next paragraph). \\
 The position of a point along the ellipse is parametrized in terms of
 the relative strong phase $\delta$.
 For both curves in the example of fig.~\ref{fig:bsin}(b), a maximal positive
 (negative) correction $\Delta \azero$ is reached for $\delta = 0 \:$
 ($\delta = \pm \pi \:$) where $\bzero = 0$.
 For not too large values of $\delta$, which correspond to $|\bzero|$
 values close to zero, the correction $\Delta \azero$ is positive and
 almost independent of $|\bzero|$. 
 On the contrary, the asymmetry correction $\Delta \azero$ decreases rapidly
 as a function of $|\bzero|$ ( see fig.~\ref{fig:bsin}(b) ) 
 when this parameter approaches
 its maximum allowed value (for a given $\alpha$ and $\beta$) and
 the corresponding $\delta$ is close to $\pm \pi/2$. For even larger
 strong phases, $\: \Delta \azero$ may become negative.
 However, as pointed out 
 previously, large values of $|\bzero|$ would only occur for unexpectedly large
 strong phases.\\
 An example where the magnitude of $\bzero$ is kept small is shown in
 fig.~\ref{fig:bsin}(c) where again we keep $\beta = 18^{\circ}$ fixed 
 and, using the solutions of (\ref{eq:sys1}) and (\ref{eq:sys2}), 
 we plot $\Delta \azero$ vs. $\azero$
 for four different values of $ \: \bzero = -0.075 \:, \:\: \: $
 $\: -0.050  \:,\:\:\:$
 $\: 0.025 \:,\:\:\: $
 $\: 0.075 \:\:\:\:$.
 In this range, we find that $\Delta \azero $ scales approximately as 
 $\cos\delta$ 
 and therefore $\Delta \azero $ is only marginally affected by $\bzero$. 
 In this example, the corresponding solutions for $\delta$ are those
 of fig.~\ref{fig:bsin}(d) where $\delta$ is plotted against the measured
 angle $\alpham$ .\\
 In conclusion, the value of the $\bzero$ parameter is found to provide
 useful information on the strong phases difference $\delta$. 
 If $\delta$ turns out 
 to assume large values ( contrary to the
 conventional expectation ), then the parameter $\bzero$ is expected to
 be large in magnitude and to contribute significantly to the assessment 
 of the ``penguin corrected''
 value of $\alpha$, which is otherwise determined by the value 
 of $\azero$ only. \\
 The current experimentally allowed range 
 for the $\bzero$ parameter is discussed in the following paragraph.\\
\begin{figure}[htbp]
  \begin{center}
      \mbox{\epsfig{figure=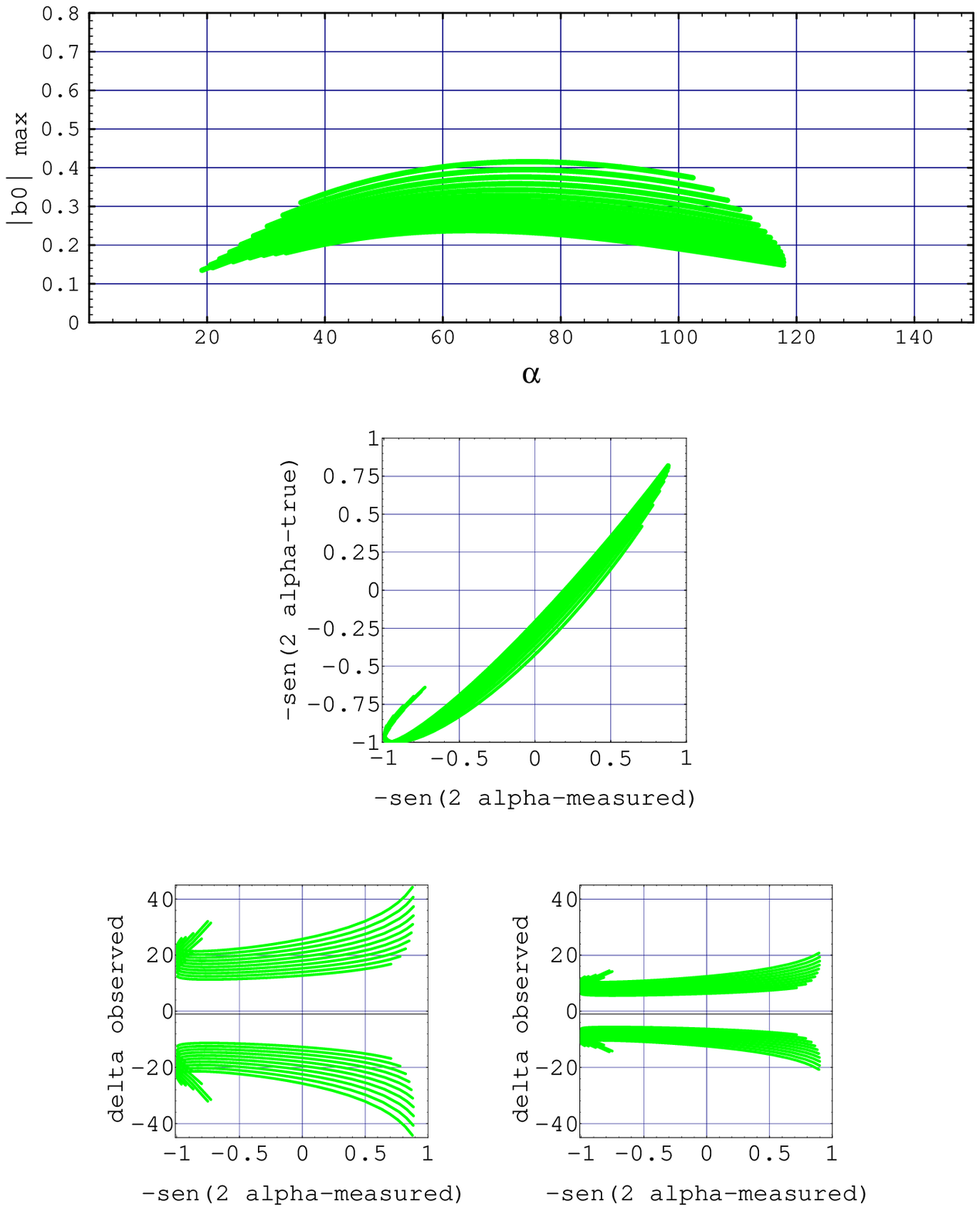,%
           height=7.0in,bbllx=103pt,bblly=120pt,bburx=596pt,bbury=623pt}}
  \end{center}
  \caption[  
 Current bounds from CKM limits on :
 (a) maximal allowed value of the $\bzero$ parameter plotted against $\alpha$ ;
 (b) penguin-corrected asymmetry vs. measured asymmetry $\azero$ ;
 (c) strong phase $\delta$ vs. measured asymmetry $\azero$ for $\bzero = \pm 0.1 $. 
 (d) same as (c) for $\bzero = \pm 0.05 \:$ .
]{\label{fig:b0limits}
 Current bounds from CKM limits on :
 (a) maximal allowed value of the $\bzero$ parameter plotted against $\alpha$ ;
 (b) penguin-corrected asymmetry vs. measured asymmetry $\azero$;
 (c) strong phase $\delta$ vs. measured asymmetry $\azero$ for $\bzero = \pm 0.1 $. 
 (d) same as (c) for $\bzero = \pm 0.05 \:$ .
           }
\end{figure}

\newpage
\subsection{ Bounds from the present CKM limits  }
\label{sec:b0bounds}

\begin{figure}[htbp]
  \begin{center}
      \mbox{\epsfig{figure=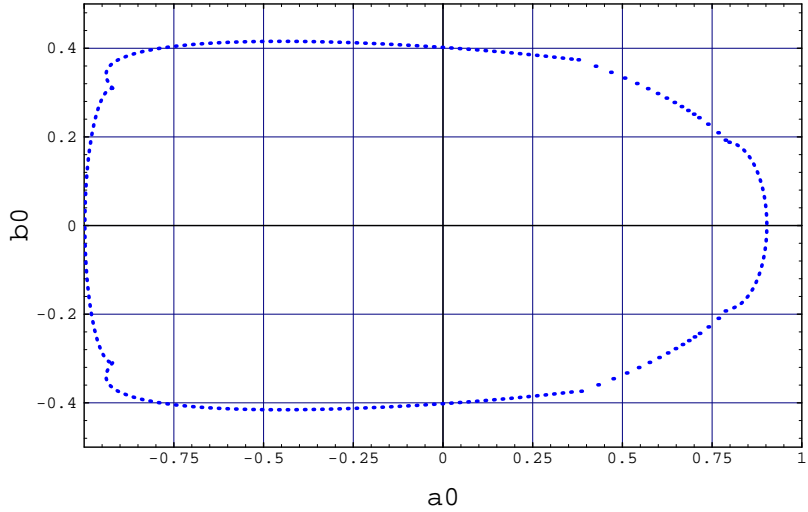,%
        height=4.5in,bbllx=83pt,bblly=140pt,bburx=576pt,bbury=643pt}}
  \end{center}
  \caption[ 
 Combined bounds for $\azero$ and  $\bzero$ 
 from the present CKM limits and with $P/T = 0.055 \: \:$.
]{\label{fig:fig8}
 Combined bounds for $\azero$ and  $\bzero$ 
 from the present CKM limits and with $P/T = 0.055 \: \:$.
           }
\end{figure}
\begin{figure}[htbp]
  \begin{center}
      \mbox{\epsfig{figure=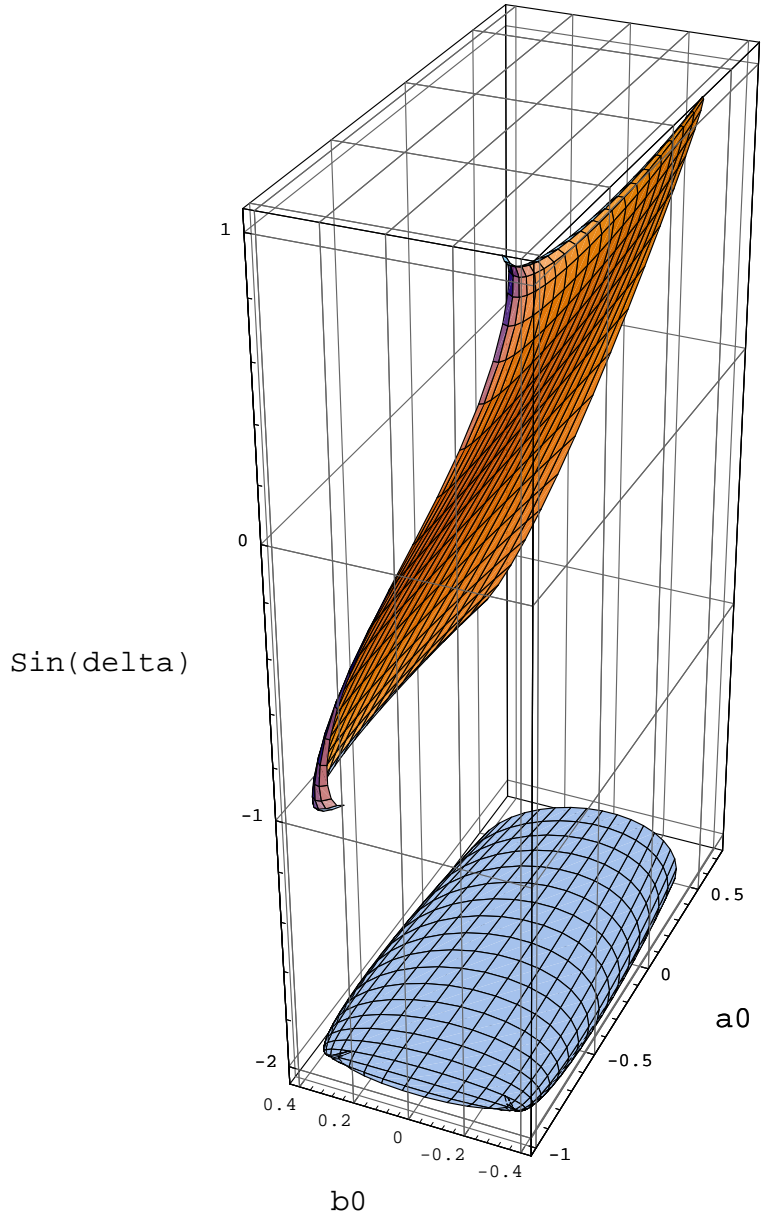,%
        height=5.0in,bbllx=103pt,bblly=70pt,bburx=596pt,bbury=573pt}}
  \end{center}
  \caption[ 
 Expected values for $\sin\delta \:$ once $\:\azero\:$, $\bzero \:$
 and $\beta$ are known. The allowed domain for ($\:\azero \:$,$\:\bzero \:$)
 from the present CKM limits and with $P/T = 0.055 \: \:$ is shown in
 the projection.
]{\label{fig:fig9}
 Expected values for $\sin\delta \:$ once $\:\azero\:$, $\bzero \:$
 and $\beta$ are known. The allowed domain for ($\:\azero \:$,$\:\bzero \:$)
 from the present CKM limits and with $P/T = 0.055 \: \:$ is shown in
 the projection.
           }
\end{figure}

 The maximum value $|\bzeromax|$ 
 allowed by the present experimental CKM limits
 \cite{bib:Ali96},\cite{bib:AliDec96},\cite{bib:Rosner96}
 is plotted in fig.~\ref{fig:b0limits}(a) as a function of 
 the true value of $\alpha$ when
 both $\alpha$ and $\beta$ are allowed to vary through their common dependence 
 on $\rho$ and $\eta$ within the domain shown in fig.~\ref{fig:bounds}(b).\\
 The values allowed at present by the experimental CKM limits
 for the ``penguin-corrected'' asymmetry $-\sin 2 \alpha$ 
 are plotted in fig.~\ref{fig:b0limits}(b)
 against the {\it measured } value of the asymmetry 
 $\azero = -2 \sin  \alpham$ 
 by varying $\: -\pi \le \delta \le \pi \:$ 
 in such a way to account for all possible allowed values of $\bzero$.
 As pointed out previously,
 the measured value of $\azero$ is significantly higher than the ``true'' value 
 of the asymmetry when $\alpham$ belongs to the
 interval $[\:45^{\circ},135^{\circ}\:]$.
 On the contrary, we can see in this picture that $\alpham$ solutions located 
 in the approximate range $[\:20^{\circ},45^{\circ}\:]$
 produce a marginal correction to the true value of $\alpha$ 
 of {\it opposite sign} with respect to the previous case 
 (i.e.: the measured value of $\azero$ is slightly smaller than 
 the ``true'' value ).\\
 Following a different approach, if we keep the magnitude of $|\bzero|\:$
 {\it fixed} and let $\alpha$ and $\beta$ vary within the present CKM bounds,
 we can extract, as a solution of (\ref{eq:sys1}) and (\ref{eq:sys2}),
 the value of $\delta$ which corresponds
 to a given observed value of the asymmetry $\azero$. This is plotted
 in fig.~\ref{fig:b0limits}(c), where 
 we have allowed for our ignorance of the sign of $\bzero$,
 and we have superimposed on the same picture two examples.
 The bounded area with positive (negative) values
 of $\delta$ corresponds to
 $\bzero = -0.1$ and $\bzero = 0.1$, respectively. 
 For large {\it negative} values of $\azero$ on the {\it left } side of 
 figs.~\ref{fig:b0limits}(c),(d) two solutions for $\delta$ are found :
 the one with larger absolute values of $\delta$ again corresponds
 to the approximate interval $[\:20^{\circ},45^{\circ}\:]$ for $\alpham$ 
 not (yet) ruled out by the present experimental limits.
 A similar example is shown in fig.~\ref{fig:b0limits}(d) for
 $\bzero = \pm 0.05 $ and where $|\delta|$ is of order $10^{\circ}\:$
 over a large fraction of the accessible range of $\azero$.\\
 Next, we want to establish the combined bounds for the $\azero$ and
 $\bzero$ parameters. 
 For a given choice of $\alpha$ and $\beta$, the 
 two parameters $\azero$ and $\bzero$ are constrained onto one ellipse 
 in the ($\azero$,$\:\bzero$) plane and the value of the
 parameter $\delta$ is used to define the position of the point
 along the curve.
 From the (exact) equations (\ref{eq:a0dev}) and (\ref{eq:b0dev}),
 the constraint turns out to be an ellipse displaced along the 
 $\azero$ axis by the amount $\azerobar$ :
\begin{equation}
 (\frac{\azero - \azerobar}{m})^2 + (\frac{\:\bzero\:}{\:\:\bzeromax\:})^2 = 1 
  \label{eq:a0b0ellisse}
\end{equation}
 where 
 $ \: \azerobar = [(\Apsuat)^2 - 1] \sin 2 \alpha \: / \: C $ and the 
 minor semi-axis is given by the expression :
 $ \: \: m = 2 \: (\Apsuat) \sin\alpha [(\Apsuat)^2 - \cos2\alpha)] \: / \: C\:$
 , while the major semi-axis $\: \: |\bzeromax| \:$ is given by :
 $(\: \: \bzeromax)^2 = 4 (\Apsuat)^2 \sin^2\alpha \: / \: C \: $
 and the common denominator is :
 $ \: \: C \approx 1 - 2 (\Apsuat)^2 \cos2\alpha $. \\
 By varying $\alpha$ and $\beta$ within their allowed domain in the
 ($\rho$,$\:\eta$) plane, a closed boundary in the ($\azero$,$\:\bzero$) plane
 is obtained which is shown in fig.~\ref{fig:fig8}.  \\
 This boundary can also be seen as the projection onto the
 ($\azero$,$\:\bzero$) plane
 of the surface represented in fig.~\ref{fig:fig9}.
 In this picture, we plot
 the value of $\sin\delta$ extracted as a solution of 
(\ref{eq:sys1}) and (\ref{eq:sys2}) for each point 
 in the allowed region of the ($\azero$,$\:\bzero$) plane
 with the assumption that the corresponding angle $\beta$,
 as well as the other two measured quantities  $\: \azero$ and  $\: \bzero$,
 are perfectly known, i.e.:  no experimental uncertainties 
 are introduced at this level.\\
 As a final remark, we want to underline that the method used 
 to derive the 
 bounds presented in this section, although applicable in general,
 depends numerically on 
 the value of $P/T = 0.055$ from section 2 
 and, in this respect, the actual values of the bounds 
 have to be considered as ``model-dependent''.

\section{Conclusions}
\label{sec:conclu}

 We have described a quantitative procedure to extract ( in a model-dependent
 way ) the value of the CKM angle $\alpha$ from a measurement of $\btopipi$
 in the presence of penguin pollution.
 We have shown that uncertainties of strong interaction origin can be
 controlled and that a model-dependent penguin correction to the
 measured value of $\alpha$ from $\btopipi$ can be evaluated and 
 parameterized in terms of the CKM angle $\beta$ and of the parameters
 $\azero$ and $\bzero$ extracted from a time-dependent analysis.
 Also, the $\cos \Delta mt \:$ term in a time-dependent
 analysis of $\bbartopipi$ is found {\it not} to be sufficient to 
 provide a direct 
 measurement of the size of the penguin, but to provide instead 
 useful information on the phases of strong origin.\\
 The model-dependence of the procedure
 stems from the numerical dependence 
 on the ratio of penguin-to-tree hadronic matrix elements $P/T$
 (see section 2)
 of the penguin correction $\Delta \alpha$ (to the weak phase $\alpha$),
 of the corresponding correction $\Delta \azero$ (to the observed asymmetry
 $\azero$) and of the ratio $\frac{\bzero}{\sin \delta}$. \\ 
 The model-dependent procedure to extract $\alpha$ and $\delta$ proposed 
 in this paper is applied
 to one final numerical example where we 
 take the value $P/T = 0.055$ from section 2 and assume for
 $\beta$ a measured central value of $\beta = 18^{\circ}$. 
 An observed asymmetry $\azero \approx -0.14$
 would, in this example, require a large penguin correction 
 $\Delta \azero \approx 0.32$ and the extracted 
``true'' asymmetry value would be $\azero^{corrected} \approx -0.46$. 
In terms of $\alpha$,
 a correction $\Delta \alpha \approx 10^{\circ}$ should be applied to 
 the measured central value $\alpham \approx 86^{\circ}$ and would result
 into a ``penguin corrected'' value of  $\alpha \approx 76^{\circ}$.
 An obvious comment is that a $\Delta \alpha $ correction of order $10 \%$
 corresponds to a significantly larger correction in terms 
 of $\Delta \azero$ as a consequence of 
 the $-\sin 2 \alpha$ dependence of the asymmetry on $\alpha$.
 In our example, a measured central value of $\bzero = -0.1$ would 
 correspond to a value for the relative strong phase  
 $\delta \approx 16^{\circ}$.\\
 The experimental errors on the parameters $\azero$
 and $\bzero$ from a time-dependent analysis of $\btopipi$, together with
the expected error from an independent measurement of $\sin 2 \beta$,
propagate into an uncertainty on the asymmetry correction $\Delta \azero$
and on the extracted value of $\delta$. \\
The evaluation of such uncertainties
as well as the assessment of the experimental  
sensitivities to $\alpha$ and $\delta$ is beyond the scope of the present
paper. 

\section*{Acknowledgments}
 We would like to thank P.Colangelo for useful discussions.



%

\newpage
\begin{thebibliography}{9}
\bibitem{bib:quinn} For a review see, e.g., 
   Y. Nir and  H.R. Quinn, {\it Annu. Rev. Nucl. Part. Sci.} {\bf 42} (1992) 
211, and references therein.
\bibitem{bib:gronau93} M. Gronau, Phys. Lett. {\bf B300} (1993) 163.
\bibitem{bib:stech} M. Bauer, B. Stech and M. Wirbel, Z. Phys. C {\bf 34} 
(1987) 103.
\bibitem{wolfes} L. Wolfenstein, Phys. Rev. D {\bf 43} (1991) 152; and  
references there.
\bibitem{bib:kramer1} G. Kramer, W.F. Palmer and Y.L. Wu, Report DESY 95-246 
(1995).
\bibitem{bib:simma} H. Simma and D. Wyler, Phys. Lett. {\bf B272} (1991) 375.
\bibitem{buras1} A.J. Buras and R. Fleischer, Phys. Lett. {\bf B341} 
(1995) 379.
\bibitem{bander} M. Bander, D. Silverman and A. Soni, Phys. Rev. Lett.
{\bf 43} (1979) 242. 
\bibitem{bib:donoghue} J.F. Donoghue, E. Golowich, A. Petrov and J.M. Soares, 
Phys. Rev. Lett. {\bf 77} (1996) 2178.
\bibitem{kamal} A.N. Kamal, Int. J. Mod. Phys. {\bf A7} (1992) 3515.
\bibitem{halperin} B. Blok and I. Halperin, Phys. Lett. {\bf B385} (1996) 
324. 
\bibitem{bib:gronau90} M. Gronau and D. London, Phys. Rev. Lett. {\bf 65} 
(1990) 3381. 
\bibitem{xing1} C. Hamzaoui and Z. Xing, Phys. Lett. {\bf B360} 
(1995) 131. 
\bibitem{he} N.G. Deshpande and X.-G. He, Phys. Lett. {\bf B384} 
(1996) 283. 
\bibitem{kramer2} G. Kramer and W.F. Palmer, Phys. Rev. D 
{\bf 52} (1995) 6411.
\bibitem{deandrea} A. Deandrea, N. Di Bartolomeo, R. Gatto, F. Feruglio 
and G. Nardulli, Phys. Lett {\bf B320} (1994) 170.
\bibitem{wolfe} J.P. Silva and L. Wolfenstein, Phys. Rev. D {\bf 49} (1994) 
1151.
\bibitem{xing2} Z. Xing, Nuovo Cim. {\bf A108} 
(1995) 1069. 
\bibitem{desh1} N.G. Deshpande and X.-G. He, Phys. Rev. Lett. {\bf 75} (1995) 
1703. 
\bibitem{hernandez} M. Gronau, O.F. Hernand\'ez, D. London and 
 J.L. Rosner, Phys. Rev. D {\bf 52} (1995) 6356.  
\bibitem{dighe} A.S. Dighe, M. Gronau and J.L. Rosner, Phys. Rev. D 
{\bf 54} (1996) 3309. 
\bibitem{mannel} R. Fleischer and T. Mannel, Report TTP 96-49 (1996).
\bibitem{bib:buras2} A.J. Buras and R. Fleischer, Phys. Lett. {\bf B360} (1995) 
138. 
\bibitem{sfigas} F. DeJongh and P. Sphicas, Phys. Rev. D {\bf 53} (1996) 4930.
\bibitem{snyder} A.E. Snyder and H.R. Quinn, Phys. Rev. D {\bf 48} (1993) 
2139;  H.J. Lipkin, Y. Nir, H.R. Quinn and A. Snyder, Phys. Rev. D {\bf 44} 
(1991) 1454.  
\bibitem{aleksan} R. Aleksan, F. Buccella, A. Le Youanc,  
L. Oliver, O. P\'ene and J.-C. Raynal, Phys. Lett. {\bf B356} (1995) 95.
\bibitem{fritz} H. Fritzsch, D. Wu and Z. Xing, Phys. Lett. {\bf B328} (1994)
477.
\bibitem{xing3} D.Du and Z. Xing, Phys. Lett. {\bf B280} 
(1991) 292, and references therein. 
\bibitem{buras3} A.J. Buras, M. Jamin, M.E. Lautenbacher and P.H. Weisz, 
Nucl. Phys. {\bf B370} (1992) 69; {\it ibid.} {\bf B400} (1993) 37; A.J. 
Buras, M. Jamin and M.E. Lautenbacher, Nucl. Phys. {\bf B400} (1993) 75.
\bibitem{ciuchini} M. Ciuchini, E. Franco, G. Martinelli and L. Reina, 
Nucl. Phys. {\bf B415} (1994) 403.
\bibitem{desh2} N.G. Deshpande and X.-G. He, Phys. Rev. Lett. {\bf 74} 
(1995) 26.
\bibitem{gronau95} M. Gronau, O.F. Hernand\'ez, D. London and 
J.L. Rosner, Phys. Rev. D {\bf 52} (1995) 6374. 
\bibitem{bauer} M. Bauer, B. Stech and M. Wirbel, Z. Phys. C {\bf 34} 
(1987) 103.
\bibitem{pdg} Review of Particle Properties, Phys. Rev. D {\bf 54} (1996) 1.
\bibitem{koide} Y. Koide, Shizuoka report US-94-05 (1994). 
\bibitem{bib:Ali96} A. Ali and D.London, Report DESY 96-140 
(1996).
\bibitem{bib:AliDec96} A. Ali, Report DESY 96-248  
(1996).
\bibitem{bib:Rosner96} J. Rosner, Report EFI-96-46 
(1996).
\end{thebibliography}
\end{document}